\documentclass[aip,pop,superscriptaddress,showpacs]{revtex4-1}

\usepackage[centertags,reqno]{amsmath}
\usepackage{amssymb}
\usepackage{graphicx} 
\usepackage{epsfig}
\usepackage{ulem}
\usepackage{eso-pic}                                                            

\def\beq{\begin{equation}}
\def\eeq{\end{equation}}
\def\beqar{\begin{eqnarray}}
\def\eeqar{\end{eqnarray}}
\def\nn{\nonumber}

\def\para{\parallel}

\newcommand{\diff}[2]{\frac{d#1}{d#2}}

\newcommand{\pdiff}[2]{\frac{\partial#1}{\partial#2}}

\newcommand{\pdt}{\partial_t}

\newcommand{\enum}[2]{{#1}\times10^{#2}} 

\newcommand{\vect}[1]{{\bf #1}}
\def\div{\nabla\cdot}
\def\grad{\nabla}

\newcommand{\gradpar}{\grad_\parallel}
\newcommand{\gradperp}{\grad_\perp}

\newcommand{\defeq}{\ensuremath{\stackrel{\text{\tiny def}}{=}}}

\newcommand{\vpar} {v_\parallel}

\newcommand{\kpar} {k_\parallel}
\newcommand{\kperp} {k_\perp }

\newcommand{\Evec}{\ensuremath{\boldsymbol{{\rm E}}}}
\newcommand{\Bvec}{\ensuremath{\boldsymbol{{\rm B}}}}
\newcommand{\Jvec}{\ensuremath{\boldsymbol{{\rm J}}}}
\newcommand{\Fvec}{\ensuremath{\boldsymbol{{\rm F}}}}
\newcommand{\fvec}{\ensuremath{\boldsymbol{{\rm f}}}}
\newcommand{\vE}{\ensuremath{\boldsymbol{{\rm v}_{E}}}}
\newcommand{\bo}{\ensuremath{\boldsymbol{{\rm b}_0}}}
\newcommand{\bvec}{\ensuremath{\boldsymbol{{\rm b}}}}
\newcommand{\xvec}{\ensuremath{\boldsymbol{{\rm x}}}}

\newcommand{\zvec}{\ensuremath{\boldsymbol{{\rm z}}}}

\newcommand{\bxgp}{\bvec\times\gradperp}

\newcommand{\vve}{\ensuremath{\boldsymbol{{\rm v}}_{e}}}
\newcommand{\vvi}{\ensuremath{\boldsymbol{{\rm v}}_{i}}}
\newcommand{\vpe}{v_{\parallel \rm e}}
\newcommand{\vpi}{v_{\parallel \rm i}}
\newcommand{\vvE}{\ensuremath{\boldsymbol{{\rm v}}_{E}}}

\newcommand{\nuei}{\nu_{ei}}
\newcommand{\nuii}{\nu_{ii}}
\newcommand{\nue}{\nu_{e}}
\newcommand{\nuen}{\nu_{en}}
\newcommand{\nuin}{\nu_{in}}

\newcommand{\rs}{\rho_{s}}
\newcommand{\ri}{\rho_{i}}
\newcommand{\wci}{\Omega_{i}}
\newcommand{\wcix}{\Omega_{ix}}

\newcommand{\tomega}{\tilde\omega}

\newcommand{\cm}{\rm cm}

\newcommand{\cmn}{{\rm cm}^{-3}}

\newcommand{\eV}{\rm eV}

\newcommand{\T}{\rm T}

\begin{document}

\title{Analysis of plasma instabilities and verification of the BOUT code
  for the Large Plasma Device}
\author{P. Popovich}
\affiliation{Department of Physics and Astronomy and Center for
  Multiscale Plasma Dynamics, University of
  California, Los Angeles, CA 90095-1547}
\author{M.V. Umansky}
\affiliation{Lawrence Livermore National Laboratory,
Livermore, CA 94550, USA}
\author{T.A. Carter}
\email{tcarter@physics.ucla.edu}
\author{B. Friedman}
\affiliation{Department of Physics and Astronomy and Center for
  Multiscale Plasma Dynamics, University of
  California, Los Angeles, CA 90095-1547}



\date{\today}

\begin{abstract}
The properties of linear instabilities in the Large Plasma Device
[W. Gekelman {\itshape et al.}, Rev. Sci. Inst., 62, 2875 (1991)] are studied both
through analytic calculations and solving numerically a system of
linearized collisional plasma fluid equations using the 3D fluid code
BOUT [M. Umansky {\itshape et al.}, Contrib. Plasma
Phys. 180, 887 (2009)], which has been successfully modified to treat cylindrical
geometry. Instability drive from plasma pressure gradients and flows is
considered, focusing on resistive drift waves, the Kelvin-Helmholtz
and rotational interchange instabilities. A general linear dispersion
relation for partially ionized collisional plasmas including these
modes is derived and analyzed. For LAPD relevant profiles including
strongly driven flows it is found that all three modes can have
comparable growth rates and frequencies. Detailed comparison with
solutions of the analytic dispersion relation demonstrates that BOUT
accurately reproduces all characteristics of linear modes in this
system.
\end{abstract}

\maketitle

\section{Introduction}

Understanding complex nonlinear phenomena in magnetized plasmas
increasingly relies on the use of numerical simulation as an enabling
tool.  The development of a robust predictive capability requires
numerical models which are verified through comparison with analytic
calculation and validated through comparison with
experiment~\cite{greenwald2010}.  A tractable analytic problem useful
for verification of numerical models of plasma turbulence and
transport is linear
stability~\cite{Naulin2008,Kasuya2007,Holland2007}.  Understanding of
linear instabilities in a set of model equations forms a framework for
developing physical insights and mathematical apparatus that can be
further used for attacking a more difficult nonlinear problem.

This paper presents a study of linear gradient-driven instabilities in
a cylindrical magnetized plasma using the Braginskii two-fluid model.
This work was undertaken with two motivations: (1) to gain
understanding of the character of linear instabilities in the Large
Plasma Device (LAPD) at UCLA~\cite{gekelman1991} and (2) to verify
linear calculations using the BOUT 3D Braginskii fluid turbulence
code~\cite{Xu1998} in cylindrical geometry.  The BOUT code was
originally developed in the late 1990s for modeling tokamak edge
plasmas; the version of the code used in this study is described in
detail by Umansky~\cite{Umansky2009}.

Instability drive in LAPD comes from plasma pressure
gradients~\cite{Penano1997} and strong azimuthal flow which can be
externally driven through biasing~\cite{Horton2005,Carter2009}.  These
free energy sources can drive resistive drift
waves~\cite{marshall1986} and Kelvin-Helmholtz and rotational
interchange instabilities~\cite{Rognlien1973}. The Kelvin-Helmholtz
instability and unstable drift-Alfv\'{e}n waves have been
experimentally observed in LAPD~\cite{Burke:2000, Maggs:2003,
Carter2009, Horton2005}. The analytic calculations and verification
runs on BOUT are performed using LAPD-like profiles of plasma density,
temperature and plasma potential (and therefore cross-field $E\times
B$ flow).  It is found that all three modes (drift waves,
Kelvin-Helmholtz, rotational interchange) can be important in LAPD
plasmas. Detailed comparison with solutions of the analytic dispersion
relation demonstrates that BOUT accurately reproduces all
characteristics of linear modes in this system. This work forms the
foundation for nonlinear modeling of turbulence and transport in LAPD,
initial results of which will be presented in a companion paper~\cite{Popovich2010b}.

This paper is organized as follows.  Section~\ref{secEquations}
introduces the LAPD geometry and presents the fluid model used for
calculations of linear instabilities.  Section~\ref{secBOUTeq}
discusses the implementation of these equations in the BOUT code,
including a discussion of techniques used to extract characteristics
of linear instabilities. Comparison of BOUT calculations to analytic
linear eigenmode solutions are presented in Section~\ref{secVerify}
for three instabilities: resistive drift waves, Kelvin-Helmholtz and
rotational interchange modes. Section~\ref{secDW_KH_IC} discusses the
linear stability of experimentally measured LAPD profiles against
these three instabilities and a discussion of the similarity to
experimental observations. The effect of ion-neutral collisions on the
linear solutions is discussed in Section~\ref{secVerif_NU}. A summary
of the paper is presented in Section~\ref{secCon}.  Appendices are
provided which cover: a derivation of the specific set of fluid
equations used in this work (Appendix~\ref{secAppendixFluid}); a
derivation of the vorticity equation used in BOUT
(Appendix~\ref{secAppendixVort}); and a list of parameters and
boundary conditions used in the verification study
(Appendix~\ref{secAppendixProfiles}).

\section{Geometry and physics model}\label{secEquations}

The geometry used in this study is that of the LAPD: a $\sim 17$~m
long cylindrical magnetized plasma with typical plasma radius
(half-width at half-maximum) of $a \sim 30{\rm~cm}$ (vacuum chamber radius
$r = 50{\rm~cm}$).  Typical plasma parameters in LAPD for a $1$~kG magnetic
field are shown in Table~\ref{tabLAPDparams}.

\begin{figure}[htbp]
\begin{center}
\includegraphics[width=0.8\textwidth]{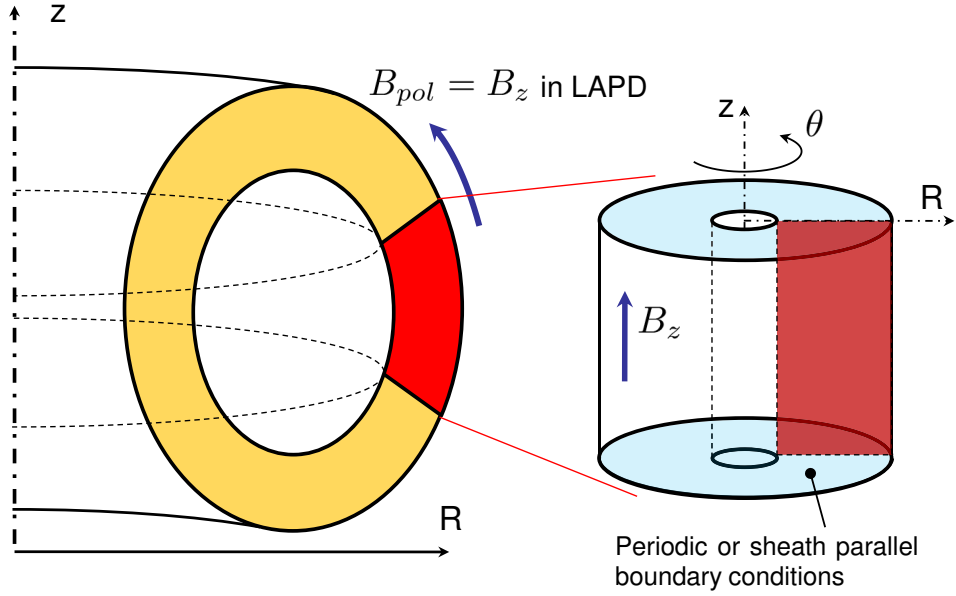}
\end{center}
\caption{\label{figGeometry} (Color Online) Schematic view of LAPD geometry
representation in the BOUT code. The poloidal direction of the tokamak
geometry becomes the axial direction $z$ in LAPD, and the toroidal
coordinate of a tokamak corresponds to the azimuthal angle $\theta$ in
LAPD.}
\end{figure}

The configuration is modeled as a cylindrical annulus to avoid the
singularity of cylindrical coordinates near the axis in the BOUT
numerical implementation (Fig~\ref{figGeometry}). Using the scheme
shown in Fig~\ref{figGeometry}, LAPD geometry can be completely
described within BOUT framework without major modification of the core
code. The only change related to geometry in the code that is
necessary is the implementation of the full cylindrical Laplacian
operator to extend the simulation domain closer to the magnetic axis.

The magnetic field is taken uniform, directed along the cylinder
axis. The axial boundary conditions are taken periodic for
simplicity. A more realistic model should include the end-plate sheath
boundary conditions, supporting potentially important wall-driven
instabilities; this will be the subject of future work.  Radial
boundary conditions used here are either zero value or zero radial
gradient.

\begin{table}
\begin{center}
  \begin{tabular}{|l|c||l|c|}
    \hline
    Species   & $^4$He            & $f_{ci}$   & $380~{\rm kHz}$ \\
    $Z$      & 1             & $\ri$    & $0.2~\cm$ \\
    $n$      & $2.5\times10^{12}~\cmn$   
                             & $\rs$    & $0.5~\cm$ \\
    $T_{\rm e}$    & $5~\eV$       & $\nuei$  & $7.4\times10^{6}~{\rm 1/s}$ \\
    $T_{\rm i}$    & $\lesssim 1~\eV$   & $\nuii$  & $5\times10^{5}~{\rm 1/s}$ \\
    $B_0$    & $0.1~\T$     & $\nuin$  & $1.2\times10^{3}~{\rm 1/s}$ \\
    $L_{||}$ & $17~{\rm m}$  & $\lambda_{ei}$  & $13~\cm$ \\
    $a$      & $\sim 0.3$~m  & $\omega_*$  & $\sim4\times10^{4}~{\rm rad/s}$ \\
    \hline
  \end{tabular}
\end{center}
\caption{\label{tabLAPDparams} Typical LAPD parameters}
\end{table}

A Braginskii two-fluid model~\cite{Braginskii1965} is used in the
analytic and BOUT calculations for instabilities in LAPD.  As evident
from Table~\ref{tabLAPDparams}, collisions are important in LAPD
plasmas: the electron collision mean free path is much smaller than
the system size parallel to the magnetic field, $\lambda_{ei} \ll
L_{||}$.  Therefore for long parallel wavelength, low frequency modes
($\omega \ll \wci$) considered here, it could be argued that the use
of a collisional fluid theory is justified.  However, it should be
noted that the quantity most important for evaluating the importance
of kinetic effects is the ratio of the parallel wave phase speed to
the thermal speed of the particles, and for drift-type modes and
Alfv\'{e}n waves in LAPD this can be near unity for the
electrons~\cite{Penano1997,Vincena2001}.  Strong collisions can disrupt
velocity-space resonant processes and it might be expected that a
fluid description becomes accurate even for $v_\phi \sim v_{\rm th}$
as $k_\parallel \lambda_{\rm ei} \rightarrow 0$, as has been shown for
ion acoustic waves through Fokker-Planck
calculations~\cite{ono1975}. The present work is part of an ongoing effort to evaluate the validity of a
fluid model (in particular that implemented in BOUT) in describing
turbulence in LAPD.  A goal of this study is to determine whether (and
how) fluid simulations can fail to describe plasma behavior, and
kinetic effects are likely to delineate when failure occurs.

The fluid equations used here represent conservation of density,
electron and ion momentum and charge:
\beqar
\label{eqfN}
\left(\pdt + \vve\cdot\grad\right)n & = &0\\
\label{eqfvpar}
\nn nm_e\left(\pdt + \vve\cdot\grad\right)\vve & = & 
-\grad p_e
-ne\left( \Evec + \frac{1}{c}\vve\times\Bvec\right)\\
& & ~~~~~~~~~~~~~ - nm_e\nuei(\vve-\vvi) - nm_e\nuen\vve\\
\label{eqfvpari}
n m_i\left(\partial_t + \vvi\cdot\grad\right)\vvi & = &
n e\left(\Evec + \frac{1}{c}\vvi\times\Bvec\right) - n m_i\nuin\vvi\\
\label{eqfdivJ}
\div\Jvec & = & 0, ~~~~~~ 
\Jvec = en(\vect{v}_{i\para}-\vect{v}_{e\para}) + 
en(\vect{v}_{i\perp}-\vect{v}_{e\perp}) 
\eeqar
where $p_e=n k_{\rm B} T_{\rm e}$. A friction term due to
ion-neutral collisions (elastic and charge-exchange) is included in
the ion momentum equation.  All terms involving finite ion temperature
effects are neglected. The friction forces in the electron momentum
equation are due to electron-ion ($\nuei$) and electron-neutral
collisions ($\nuen$). However, as Coulomb collisions are dominant for
the electrons ($\nuei \gg \nuen$), electron-neutral
collisions are ignored.

The following simplifying assumptions are made, which are
relevant for LAPD plasma parameters: constant magnetic field $\Bvec =
B_0\zvec$, $\vpe\gg\vpi$, $T_e\gg T_i$, and
no background parallel flows.  In addition, it is assumed that the
instabilities do not generate perturbations in the electron temperature.
Throughout the paper plasma density, temperature and magnetic field
are normalized to reference values $n_x$, $T_{ex}$ (chosen as the
maximum of the corresponding equilibrium profiles), and $B_0$, the
axial magnetic field. Frequencies and time derivatives are normalized
to $\wcix = eB_0/m_ic$: $\hat{\pdt} = \pdt/\wcix$, $\hat{\omega} =
\omega/\wcix$; velocities are normalized to the ion sound speed
$C_{sx}=\sqrt{T_{ex}/m_i}$; lengths -- to the ion sound gyroradius
$\rho_{sx} = C_{sx}/\wcix$; electrostatic potential to the reference
electron temperature: $\hat{\phi} = e\phi/T_{ex}$.  Further the
$''~\hat{~}~''$ symbol for dimensionless quantities will be dropped
for brevity of notation.

Combining Eqs.(\ref{eqfN}-\ref{eqfdivJ}) and linearizing (see
Appendix~\ref{secAppendixFluid}), one obtains:
\beqar\label{eqs_orig}
\nn
\partial_t N +
\bo\times\gradperp\phi_0\cdot\grad N & = & -
\bo\times\gradperp\phi\cdot\grad N_0 - N_0\gradpar\vpe\\
\partial_t{\vpe}+\bo\times\gradperp\phi_0\cdot\grad\vpe & = &
-\mu\frac{T_{e0}}{N_0}\gradpar N + \mu\gradpar\phi - \nue\vpe\\
\nn
N_0\gradpar\vpe & = & -\gradperp\cdot
       \left(
             N_0\partial_t\gradperp\phi + \partial_t N\gradperp\phi_0\right.\\
\nn& &\hspace{2.0cm} + \bo\times\gradperp\phi_0\cdot\grad\left(N_0\gradperp\phi_0\right)\\
\nn
& &\hspace{2.0cm} + \bo\times\gradperp\phi_0\cdot\grad\left(N_0\gradperp\phi\right)\vphantom{\frac{}{1}}\\
\nn
& &\hspace{2.0cm} + \bo\times\gradperp\phi\cdot\grad\left(N_0\gradperp\phi_0\right)\vphantom{\frac{}{1}}\\
\nn
& &\hspace{2.0cm} + \bo\times\gradperp\phi_0\cdot\grad\left(N\gradperp\phi_0\right)\vphantom{\frac{}{1}}\\
\nn
& &               + N_0\nuin\gradperp\phi_0
                  + N_0\nuin\gradperp\phi
                  + N\nuin\gradperp\phi_0
\left. \vphantom{\pdiff{1}{t}}\right), 
\eeqar
where $N_0, \phi_0, T_{e0}$ are zero-order (equilibrium) quantities
and $N, \phi, \vpe$, are first order perturbations; $\mu=m_i/m_e$.

Note that Eqs. (\ref{eqs_orig}) contain a zero-order term,
$\gradperp\cdot\left(\nuin N_0\gradperp\phi_0\right)$, which restricts
the choice of background profiles in the presence of neutrals. If this
term is not zero for a particular choice of $N_0(r)$ and $\phi_0(r)$
functions, then the plasma is not in mechanical equilibrium. In such a
case, an extra zero-order force, e.g., from externally applied radial
electric field, should be added to the momentum equation to balance
the force of friction with the neutrals that slows down the plasma
rotation.

Next, Eqs.~(\ref{eqs_orig}) are projected onto cylindrical coordinate
system $(r,\theta,z)$. Solutions of the form
$f(\vect{x})=f(r)\exp(im_\theta\theta + i\kpar z - i\omega t)$ are sought, where
$\kpar=2\pi n_z/L_\parallel$, $n_z$ is the parallel mode
number. Denoting $f'=\partial_r f$ and introducing the Doppler-shifted
frequency $\tomega=\omega - \frac{m_\theta}{r}\phi_0'$, the 1D
equation for radial eigenfunctions of the perturbed potential
$\phi(r)$ can be written:
\beq\label{eqPhi1D}
C_2(r)\phi'' + C_1(r)\phi' +C_0(r)\phi = 0,
\eeq
where the coefficients $C_i(r)$ are functions of equilibrium
quantities and of $\tomega$ (full expressions for $C_i$ are presented
in Appendix~\ref{secAppendixFluid}).

Equation (\ref{eqPhi1D}) is a 2nd order ordinary differential equation
(ODE) in $r$. Supplemented with proper boundary conditions on the radial
boundaries it forms a well-posed eigenvalue problem. In general
Eq. (\ref{eqPhi1D}) has to be solved numerically, due to the complex
form of the coefficients $C_i$. Note that although Eq. (\ref{eqPhi1D})
is better suited for theoretical analysis than the original system,
Eqs. (\ref{eqs_orig}), a complication for practical numerical solution
of Eq. (\ref{eqPhi1D}) is that the eigenvalue $\omega$ enters
nonlinearly the coefficients $C_i$. Therefore, a numerical solution
for the eigenvalues is easier to carry out using the original system
Eqs. (\ref{eqs_orig}), which can be cast to a standard linear algebra
eigenvalue problem amenable to solution by a standard eigenvalue
package.


\section{Solving by time-evolution with BOUT}\label{secBOUTeq}

The present version of the BOUT code~\cite{Umansky2009} is a rather
general framework suitable for integration of a system of
time-evolution PDEs in 3D space of the form
$\pdt{\fvec}=\Fvec(\fvec,\xvec)$, where the right-hand-side $\Fvec$
contains a combination of spatial differential operators applied to
the state vector $\fvec$. The right-hand-side $\Fvec$ is discretized
on a spatial mesh by finite-differencing, which results in a system of
ODEs that are integrated in time by an implicit ODE solver package
PVODE~\cite{pvode}.

For the calculations presented here, the following set of
equations are used in BOUT which are equivalent to
Eqs.~(\ref{eqfN}-\ref{eqfdivJ}):
\beqar
\label{eqBN}
\pdt N & = & - \vE\cdot\grad N - \gradpar(\vpe N)\\
\label{eqBvpar}
\pdt\vpe & = & - \vE\cdot\grad \vpe -\mu\frac{T_{e0}}{N_0}\gradpar N + \mu\gradpar\phi - \nue\vpe\\
\label{eqVort}
\pdt\varpi & = &
-\vE\cdot\grad\varpi 
- \gradpar(N\vpe) 
+ \bvec\times\grad N\cdot\grad{v_E^2}/2
-\nuin\varpi
\eeqar
where the potential vorticity 
\beq\label{eqDefVort}
\varpi\defeq\gradperp\cdot\left(N\gradperp\phi\right) 
\eeq
is introduced. While the variables $N$, $\vpe$ and $\varpi$ are
advanced in time, Eq. (\ref{eqDefVort}) is inverted on each
evaluation of the right-hand side of Eqs. (\ref{eqBN}-\ref{eqVort}) to reconstruct
the perturbed potential $\phi$ from $\varpi$.

The vorticity evolution equation, Eq. (\ref{eqVort}), replaces the
current continuity equation (\ref{eqfdivJ}) in BOUT.  Derivation of
this form of the vorticity equation from Eq.~(\ref{eqfdivJ}) is
presented in Appendix \ref{secAppendixVort}. Note that
Eq. (\ref{eqVort}) is equivalent to expression (76) in Simakov and
Catto~\cite{Simakov2003}, apart from the ion-neutral collision term
which is included in this work and all the terms involving ion
temperature which are neglected in the present work.  The third term in
the right-hand side of Eq.~(\ref{eqVort}) is important even in the
linear regime if strong background flows are present. Thus, it is
essential in both linear and nonlinear simulations of LAPD
experiments, which typically have strong azimuthal flows on the order
of Mach number M $\sim$ 0.2 for spontaneous flows and M $\sim$ 1 in
bias-induced rotation experiments \cite{Carter2009}.

To compare BOUT solution of the initial-value problem with the direct
solution of eigenvalue problem corresponding to the discretized
dispersion relation (\ref{eqs_orig}), the equations are linearized
(Eqs.~\ref{eqBN}-\ref{eqDefVort}) and advanced in time using BOUT
from a small initial seed perturbation. BOUT produces
perturbations, in this case, of density and vorticity/potential, as
functions of 3D space and time.

A specific azimuthal mode number $m_\theta$ is selected by Fourier
filtering in the azimuthal angle during the BOUT simulation. The parallel
wave number $\kpar$ is set by the length of the device and the
periodic boundary conditions in the parallel direction. 

The radial form of the numerical solution is dominated by the fastest
growing radial eigenmode. Once the solution ``locks in'' to the
fastest mode, we calculate the growth rate by fitting the time
evolution of the volume-averaged amplitude of potential fluctuations
to an exponential. The frequency of the mode is then calculated by
fitting the perturbed potential (with the exponential growth factored
out) with a sine wave at each spatial position.

\section{Verification of BOUT against eigenvalue solution}\label{secVerify}

\subsection{Electrostatic resistive drift wave}\label{secVerif_DW}

In the absence of strong flows, the resistive drift mode is likely to
be the primary instability in LAPD. In this section, the BOUT solution
in LAPD geometry is verified using a reduced subset of fluid equations
Eqs.~(\ref{eqBN}-\ref{eqDefVort}) which support only the resistive
drift instability branch.

The simplest model of the resistive drift wave can be written as a
subset of the system Eqs.~(\ref{eqBN}-\ref{eqVort}):
\beqar
\nn\pdt N & = & - \vvE\cdot\grad N_0\\
\pdt\vpe & = & -\mu\frac{T_{e0}}{N_0}\gradpar N + \mu\gradpar\phi - \nue\vpe\\
\nn\pdt\varpi & = & - \gradpar(N\vpe)
\eeqar

These equations can be combined together to form a well-known local
dispersion relation~\cite{Chen1984} that assumes 1D dependence
of the background density with constant gradient length
$L_n=N_0(x)/N_0'(x)$:
\beq\label{e:driftmode_simple}
\left(\frac{\omega}{\omega_*}-1\right)i\frac{\sigma_\para}{\omega_*}
 + \left(\frac{\omega}{\omega_*}\right)^2 = 0,
\eeq
where $\displaystyle{
\omega_*=\frac{k_\perp}{L_n}\frac{T_{e0}}{m_i\Omega_i}
}$, 
$\displaystyle{
\sigma_\para=\frac{\Omega_i\Omega_e}{\nu_{ei}}\frac{k_\para^2}{k_\perp^2}
}$.

\medskip

BOUT calculations were first verified on this simple local solution,
Eq. (\ref{e:driftmode_simple}), finding good agreement for a range
of plasma parameters. Due to its simplicity, this solution provides useful
insight into the behavior of the growth rates and frequencies. In a
bounded plasma, the dispersion relation Eq. (\ref{e:driftmode_simple})
together with a set of boundary conditions yields a set of discrete
linearly unstable modes. Among these discrete modes, the fastest
growing one is the mode that corresponds to dimensionless parameter
$\sigma_\para/\omega_*$ closest to 1.

\begin{figure}[htbp]
\begin{center}
\includegraphics[width=1.0\textwidth]{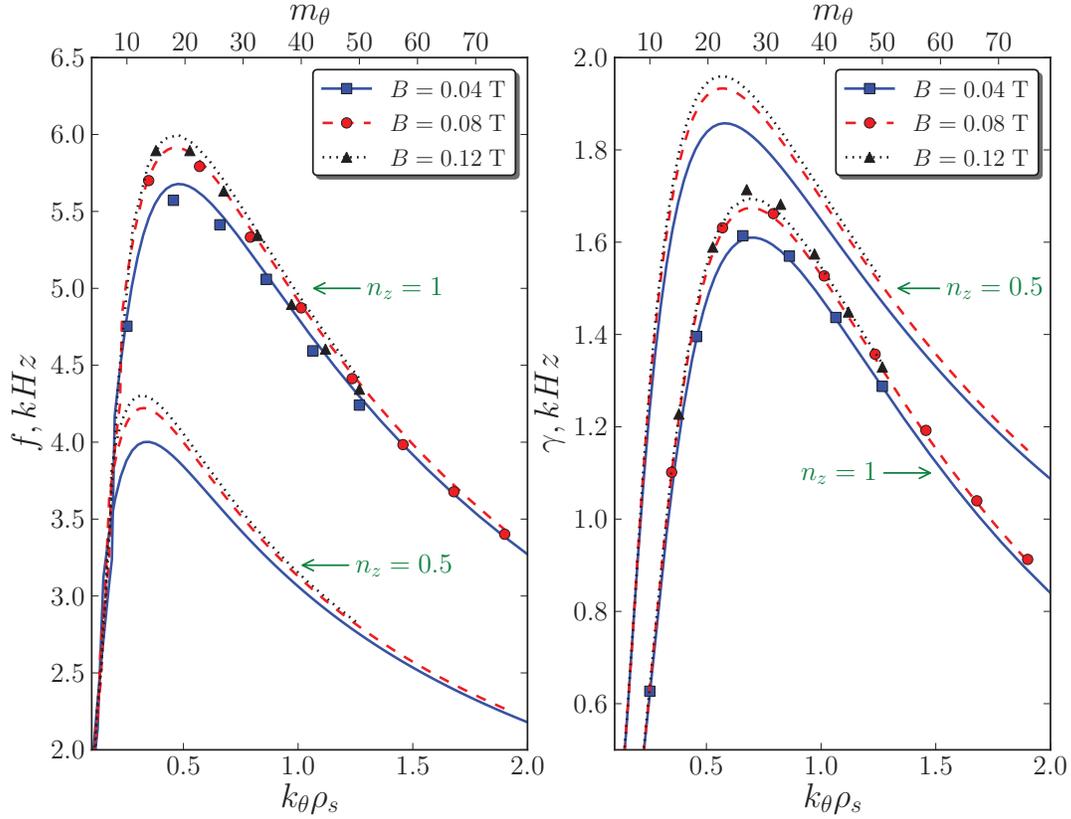}
\end{center}
\caption{\label{figDW_Bscan} (Color Online) Frequencies (left) and
  growth rates (right) of the resistive drift wave in LAPD
  configuration for experimental density profile, without equilibrium
  flows.  Analytic results are shown with lines for $B=0.04~\T$
  (solid), $0.08~\T$ (dash), $0.12~\T$ (dotted); corresponding
  BOUT results are shown with squares, circles and triangles. The two
  groups of lines correspond to the axial harmonics $n_z=1$ and
  $n_z=0.5$.  $m_\theta$ values on top axis are given for $B=0.04\T$.}
\end{figure}

The next, and more interesting, step is to compare BOUT calculations
to the eigenvalue solution of the full non-local drift wave problem
Eqs.~(\ref{eqfN}-\ref{eqfdivJ}). Here all terms in
Eqs.~(\ref{eqBN}-\ref{eqDefVort}) are retained. There is no
background potential $\phi_0$ for these calculations which eliminates
the Kelvin-Helmholtz and the rotation-driven interchange
instabilities, and only allows for the drift wave solution. There is
no simple analytic dispersion relation in this case. For comparison,
we are using the direct numerical solution of the linear problem
Eq.~(\ref{eqs_orig}) obtained with an eigenvalue solver, as described
in Section~\ref{secEquations}. The results of this comparisons for
  cylindrical geometry with relevant to LAPD parameters and profiles
  are presented in Fig.~(\ref{figDW_Bscan}). BOUT recovers the
  frequencies and growth rates for a range of magnetic field values
  ($B=0.04$, $0.08$, and $0.12$T). There is one-to-one correspondence
  between the eigenvalues found by the analytic solver and the BOUT
  solution.  Typically, the discrepancy between the two methods is
  less than 2\% for radial grids of 50 points, and the results
  converge with grid size.  For comparison, frequencies and growth
  rates for longer wavelengths, $n_z=0.5$ (fundamental mode), are also
  shown (dashed lines).

\begin{figure}[htbp]
\begin{center}
\includegraphics[width=0.9\textwidth]{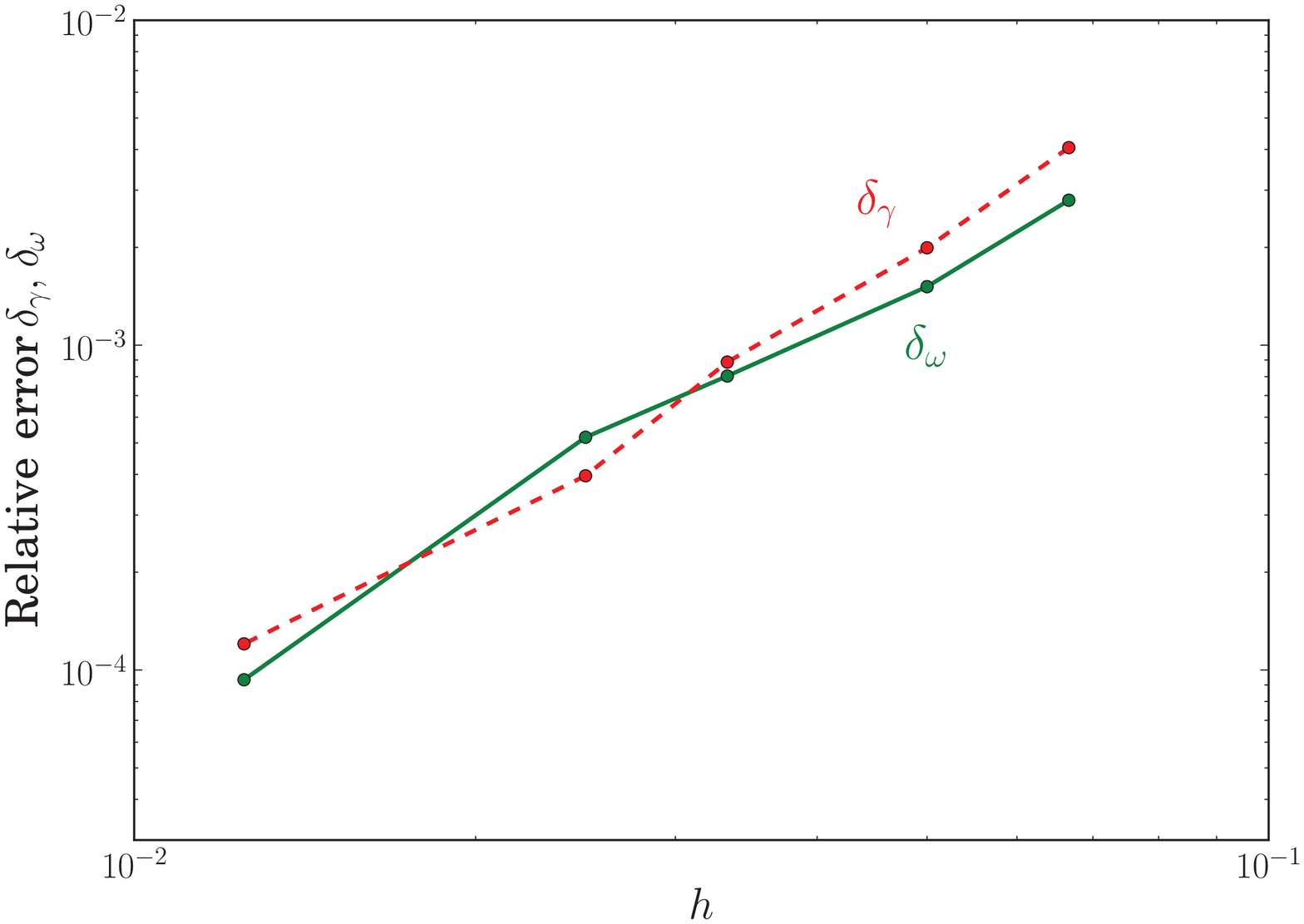}
\end{center}
\caption{\label{figConverge} (Color Online) Relative error of the growth rate and
frequency determined from the initial-value simulation as a function
of the radial grid size, indicating 2nd order convergence.}
\end{figure}

As an example of numerical convergence, the relative error of the
growth rate and frequency as a function of radial grid size $h=1/N_r$
is shown in Fig.~\ref{figConverge}. The relative error is defined here
as the difference between BOUT solution and the projected value at
$h=0$, $\delta_\gamma=|\gamma-\gamma_{h=0}|/\gamma_{h=0}$, and
analogously for the frequency $\omega$. The growth rate and frequency
extracted from the initial-value simulation converge approximately
quadratically in $h$. The difference between the BOUT solution at $h\to 0$
and the eigenvalue solver result is $0.43\%$ for the frequency and
$0.23\%$ for the growth rate. This residual error is due to the
limited numerical resolution in the azimuthal and parallel directions
(both remain fixed at 16 grid points for this convergence study) and
slight differences in the representation of the underlying equilibrium
in BOUT and the eigenvalue solver.


\subsection{Kelvin-Helmholtz instability}\label{secVerif_KH}

LAPD plasmas often involve large azimuthal flows, especially with
biasing of the vacuum vessel wall~\cite{Maggs2007, Carter2009}. The flows in the
experiments with externally applied radial bias can reach Mach number
of about 1, or $v_\theta\sim 10^6{\rm~cm/s}$. These
speeds are much higher than the typical phase velocity of the drift
wave, $v_d\sim 0.5\times10^4{\rm~cm/s}$. Also, the growth rates of the
instabilities generated by bias-induced flows can be comparable to
that of the drift wave (see Section~\ref{secDW_KH_IC}), therefore it
is essential to include these flows in the model.

Kelvin-Helmholtz (KH) instability, driven by sheared flows, represents
an interesting case for BOUT simulations in LAPD geometry, and
provides a test of the implementation of the terms involving $\phi_0$ in
BOUT. Observations of KH instability in LAPD plasmas have previously
been reported by Horton et al.~\cite{Horton2005}.

A simple model for the KH instability can be obtained from the charge
conservation equation, Eq.~(\ref{eqfdivJ}). Assuming no variation of
equilibrium or perturbed quantities along the magnetic field (flute
modes), only the polarization current contributes to this equation:
\beq
\Jvec_\perp = en(\vect{v}_{i\perp}-\vect{v}_{e\perp}) 
= -\frac{c^2m_in}{B^2}~\diff{\gradperp\phi}{t}
= -\frac{c^2m_in}{B^2}\left(\pdt + \vE\cdot\grad\right)\gradperp\phi
\eeq
For simplicity, the case of constant plasma density and
magnetic field is considered. The charge conservation equation can then be written
as
\beq\label{eqKHphi} \left(\pdt +
\vE\cdot\grad\right)\gradperp^2\phi=0 \eeq
Linearizing Eq.~(\ref{eqKHphi}) for slab geometry with periodic
coordinate $y$ we obtain the eigenvalue equation~\cite{Ganguli1997,Horton2005}:
\beq\label{eqphi}
\phi''(x) - \left(k_y^2 + \frac{k_y\phi_0'''(x)}{k_y\phi_0' - \omega}   \right)\phi = 0
\eeq
where the solution is assumed of the form $\phi(\vect{r},t) =
\phi(x)\exp(ik_yy-i\omega t)$

Analytic solution of this equation can be found for a specific choice
of stream function $\phi_0$ by matching $\phi$ and its derivative jump
at the points of singularity:
\beq\label{phi0delta}
\phi_0(x) = 
\left\{
\begin{array}{lr}
  0, & x\le -1\\
  x^2/2+x+1/2, & -1\le x\le 0\\
  -x^2/2+x+1/2, & 0\le x\le 1\\
  1, & x > 1
\end{array}
\right.
\eeq

For direct comparison with BOUT, a solution must be found with
boundary conditions imposed on a finite interval. 
We consider boundary conditions $\phi(-2)=\phi(2)=const$. In this
case, the eigenvalues are
\beq
\omega = (e^{2k_y}-1)/(2+2e^{2k_y})
\eeq
for the neutrally stable branch, and
\beqar
\nn\omega & = & \left(1-e^{4k_y}+2k_y+2k_ye^{4k_y}~\pm~\sqrt{G(k_y)}\right)/(4 + 4e^{4k_y}), \\
\nn G(k_y) & = & 9-16e^{2k_y}+14e^{4k_y}-16e^{6k_y}+9e^{8k_y} + 12k_y 
-12k_ye^{8k_y} \\
\nn & & + 4k_y^2 + 8k_y^2e^{4k_y} + 4k_y^2e^{8k_y}.
\eeqar
for the stable/unstable branches. One of the branches is unstable for
$0\le k_y\le 1.815$, maximum growth rate is 0.2346 at $k_y\approx
1.241$. This result is similar to the calculation presented by Horton
et al. for slightly different boundary conditions~\cite{Horton2005}.

\begin{figure}[htbp]
\begin{center}
\includegraphics[width=0.8\textwidth]{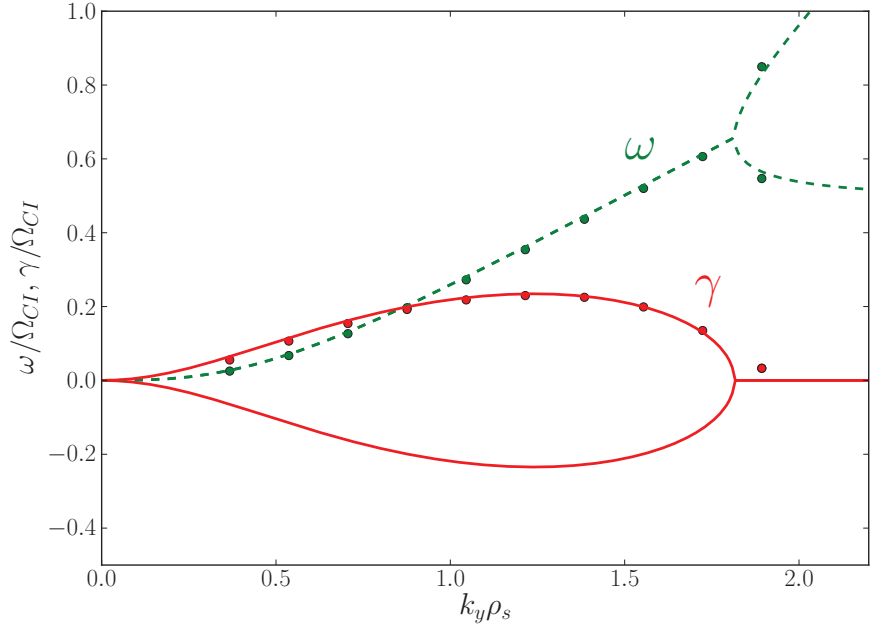}
\end{center}
\caption{\label{figKHdelta} (Color Online) Analytic solution (frequency -- dashed
    line, growth rate -- solid line) and BOUT simulations (circles)
    for the Kelvin-Helmholtz instability in slab geometry. $\phi_0$
    profile for this case is given by Eq.~(\ref{phi0delta}).}
\end{figure}

This instability is found with BOUT by solving Eq.~(\ref{eqKHphi})
written in terms of vorticity:
\beq\label{eqKHvort}
\pdt\varpi + \vE\cdot\nabla\varpi = 0
\eeq
Eq. (\ref{eqKHvort}) is explicitly linearized in BOUT and solved in
slab geometry with the same boundary conditions
$\phi(-2)=\phi(2)=const$. In BOUT, slab geometry is approximated as a
small azimuthal segment of a large aspect ratio thin annulus. The
exponential growth rate and the mode frequency is extracted from the
time evolution of the perturbed potential $\phi$. Using this method,
the frequencies and growth rates of the direct eigenvalue
solution are recovered, as shown in Fig.~\ref{figKHdelta}.

Note that the third derivative of $\phi_0$ that is present in
Eq.~(\ref{eqKHvort}) (as can be seen from Eq.~(\ref{eqphi})) is
singular, but it does not directly enter BOUT equations. BOUT uses
$\phi_0$ profile as input, which is a smooth function. Therefore, the
code has no difficulty reproducing the analytic solution even though
it implies a singularity in the $\phi_0'''$ profile.

\begin{figure}[htbp]
\begin{center}
\includegraphics[width=0.8\textwidth]{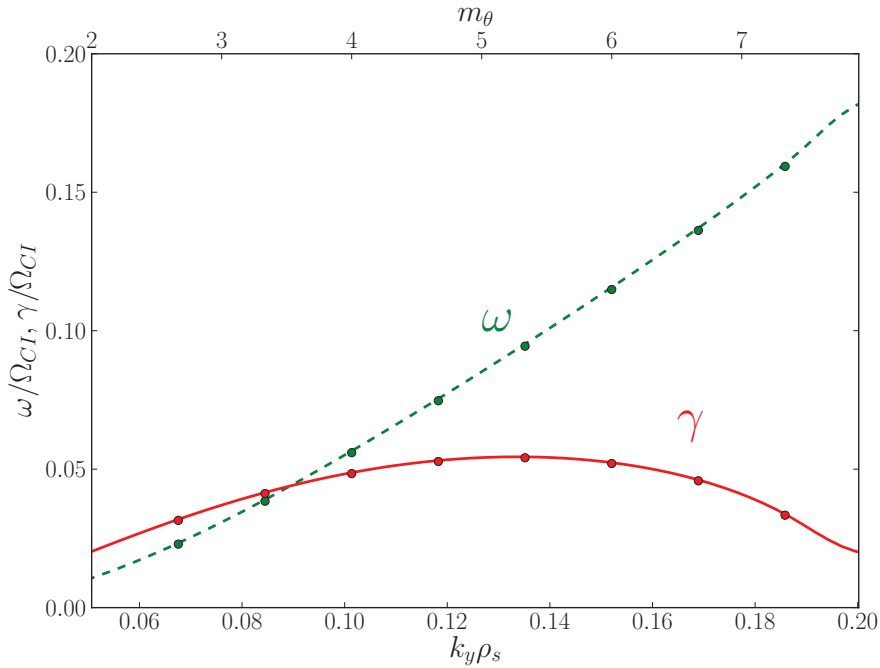}
\end{center}
\caption{\label{figKHgauss} (Color Online) Eigenvalues (frequency -- dashed line,
    growth rate -- solid line) of the Kelvin-Helmholtz instability
    as a function of perpendicular wavenumber. Circles -- BOUT
    results. Cylindrical geometry, experimental density profile, LAPD
    plasma parameters.}
\end{figure}

Next, to make a calculation relevant to the experiment, the KH
instability in LAPD geometry is considered using the experimental
density profile and a model $\phi_0(r)$ profile with amplitude values
relevant to the experiment. The background potential profile is
similar to expression (\ref{phi0delta}), but the delta-functions
in $\phi_0'''$ are replaced by Gaussians (exact expression is given in
the Appendix~\ref{secAppendixProfiles}, Eq.~(\ref{eqPhi0Gauss})).
This calculation represents a strong test of the terms involving
background flows in Eqs.~(\ref{eqVort}-\ref{eqDefVort}) since some of
these terms only contribute when both $\grad N_0$ and $\grad\phi_0$
exist. Note also that with non-constant $N_0(r)$, the density perturbation
is not zero, unlike in the situation considered above. There is no
analytic solution in this case, therefore we compare the BOUT solution
with the results of the eigenvalue solver for the system of equations
(\ref{eqfN}-\ref{eqfdivJ}). The comparison is presented in
Fig.~\ref{figKHgauss}. The result is similar to the previous KH case,
with a cutoff in perpendicular wavenumber. In LAPD geometry,
for this particular choice of profiles, this cutoff translates into
$m_\theta\approx 8$; the KH mode is stable above this value.  BOUT
reproduces the direct eigenvalue solution with a very good accuracy of
$\lesssim 2$\% for a 100 point radial grid size.

\subsection{Interchange instability}\label{secVerif_IC}

Strong azimuthal flows in LAPD not only affect the frequency of the
waves through a Doppler shift, but can also modify the growth rate even
for uniform rotation due to the induced centrifugal force. In this
section the rotation-driven interchange mode is considered
in the presence of background density gradient.

To separate the interchange mode from the other instabilities, 
the parallel wave number is set to zero (this removes the drift wave branch
of the dispersion relation) and a uniform rotation profile is chosen with
normalized rotation frequency $\Omega$: $\phi_0 = \phi_x
r^2/r_{max}^2=\Omega r^2/2$ (this removes the KH mode). Assuming an
exponential equilibrium density profile with gradient scale length $L_n$,
Eq.~(\ref{eqPhi1D}) is written as
\beq\label{eqPhiIC1}
\phi'' + \frac{1}{r}\phi' - \frac{1}{L_n}\phi'
- \frac{1}{rL_n}\left(\frac{d^2}{\tomega^2}+\frac{2d}{\tomega}\right)\phi
-\frac{m^2}{r^2}\phi = 0,
\eeq
where $d$ is the Doppler shift, $d=m\phi_0'/r=2m\phi_x/r_{max}^2=m\Omega$,
$\tomega=\omega-d$, and $L_n=N_0/N'_0$.

There are two tractable limits where analytic solution can be found,
$k L_n \gg$ 1 (slow variation), and $k L_n \ll$ 1 (sharp interface).

For small density gradient, the $1/L_n$ term can be dropped when
compared with the $1/r$ term. Employing a change of variable
$x=\sqrt{r}$, Eq.~(\ref{eqPhiIC1}) is rewritten as Bessel's equation:
\beq\label{eqPhiIC2}
x^2\phi''(x) + x\phi'(x) + \left(4C^2 - 4m^2x^2\right)\phi(x) = 0
\eeq
where $\displaystyle{C^2 =
-\frac{1}{L_n}\left(\frac{d^2}{\tomega^2}+\frac{2d}{\tomega}\right)}$. The
solution is given as a sum of Bessel functions of the first and the
second kind:
\beq\label{eqPhiIC3}
\phi(x) = C_1 J_{2m}(-2Cx) + C_2 Y_{2m}(-2Cx).
\eeq
The dispersion relation is obtained by imposing the boundary
conditions $\phi(r_{min})=\phi(r_{max})=0$ on this function. For
simplicity $r_{min}=0$ is chosen. $Y_{2m}(x)$ diverges at the
axis, so the dispersion relation in this case is given by the
condition
\beq\label{eqPhiIC4}
J_{2m}(-2C\sqrt{r_{max}})=0.
\eeq
For large $m_{\theta}$ the position of the first zero of the Bessel
function $J_m(x)$ can be estimated~\cite{AbramowitzStegun}
 as $m_{\theta}$ (e.g. $j_{m}=36.1$
for $m=30$ and the relative error monotonically decreases for larger
$m_{\theta}$). This results in a simple
approximate equation for the interchange eigenmode:
\beq\label{eqPhiIC5}
-C\sqrt{r_{max}}=m,
\eeq
which yields the approximate dispersion relation (again using $L_n\gg r_{max}$)
\beq\label{eqICdispLoLn}
\omega = m\Omega \pm i\Omega\sqrt{\frac{r_{max}}{L_n}}
\eeq
Note that the growth rate
$\gamma=\Omega\sqrt{r_{max}/L_n}$ can be
obtained from the well known dispersion relation of the
Rayleigh-Taylor instability driven by gravity, $\gamma\sim
\sqrt{g/L_n}$, if gravity is replaced by the centrifugal force of the rotation, 
$g=\Omega^2r_{max}$.
%

\begin{figure}
\begin{center}
\includegraphics[width=0.9\textwidth]{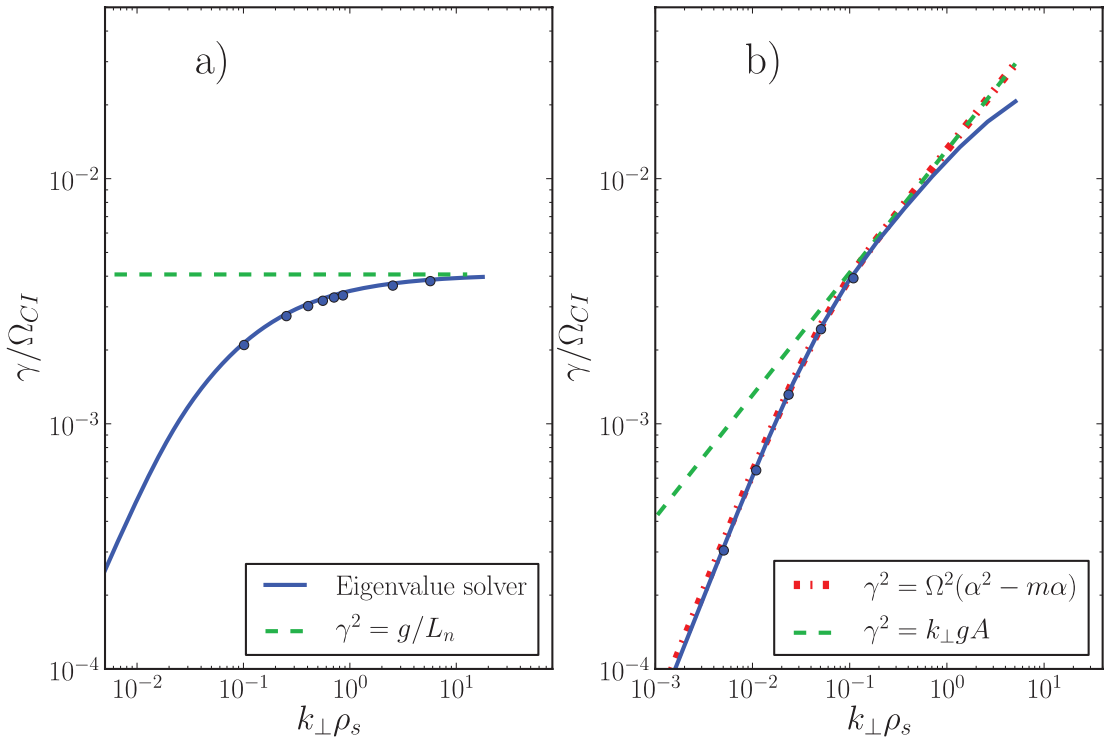}
\end{center}
\caption{\label{figIC_analytic} (Color Online) Interchange mode in a rotating
    cylinder for the case of exponential density profile (a) and
    piecewise-constant density (b). Solid line -- eigenvalue solution,
    dashed line -- asymptotic at large $k_\perp$, dashdot line --
    exact analytic solution for case (b), circles -- BOUT results.}
\end{figure}

The growth rate given by Eq.~(\ref{eqICdispLoLn}) is independent of
$m_{\theta}$ and represents an asymptotic solution for large
$m_{\theta}$. This asymptotic solution and the
exact solution of Eq.~(\ref{eqPhiIC1}) are shown in
Fig.~\ref{figIC_analytic}(a).

Another limit where a simple analytic solution can be found is the
case of a piecewise-constant density profile with a sharp interface,
$N_0(r)=N_1$ for $r<r_0$ and $N_0(r)=N_2$ for $r>r_0$,
$r_0=r_{max}/2$, $N_1>N_2$. Eq.~(\ref{eqPhiIC1}) at $r\ne r_0$ then
becomes
\beq\label{eqPhiIC6}
\phi'' + \frac{1}{r}\phi'-\frac{m^2}{r^2}\phi = 0,
\eeq
with the general solution $\phi(r)\sim r^{\pm m}$. Matching the values
of $\phi(r)$ at the interface, applying the boundary condition at the
conducting shell $\phi(r_{max})=0$, and integrating
Eq.~(\ref{eqPhiIC1}) in a small region near the interface to account
for the jump in $N_0$ and $\phi'$, we obtain the dispersion relation
\beq\label{eqICpwc}  
\tomega = \Omega\left(-\alpha \pm \sqrt{\alpha^2 - m\alpha}\right),
\eeq
where $\alpha = A(2^{2m}-1)/(2^{2m}-A)$ and $A$ is the Atwood number
$A=(N_1-N_2)/(N_1+N_2)$.

In the limit of large $m_{\theta}$ cylindrical effects become
insignificant and the growth rate converges to that of the
gravity-driven Rayleigh-Taylor instability in a slab for two fluids
with sharp interface, $\gamma\approx
\sqrt{\Omega^2mA}=\sqrt{kgA}$. The exact solution and the asymptotic solution
at large $m_{\theta}$ are shown in Fig.~\ref{figIC_analytic}(b). The
solid line represents the eigenvalue solution of the system
(\ref{eqs_orig}) where the piecewise-constant density profile is
approximated with $\tanh$ function. At higher $m_{\theta}$ numbers
(here, at $m\gtrsim 20$, or $k_\theta\rho_s\gtrsim 1$) the finite
width of the interface region becomes important compared to
$1/k_\theta$, so the numerical (eigenvalue) solution starts to deviate
from the analytic solution (\ref{eqICpwc}).

\begin{figure}[htbp]
\begin{center}
\includegraphics[width=0.9\textwidth]{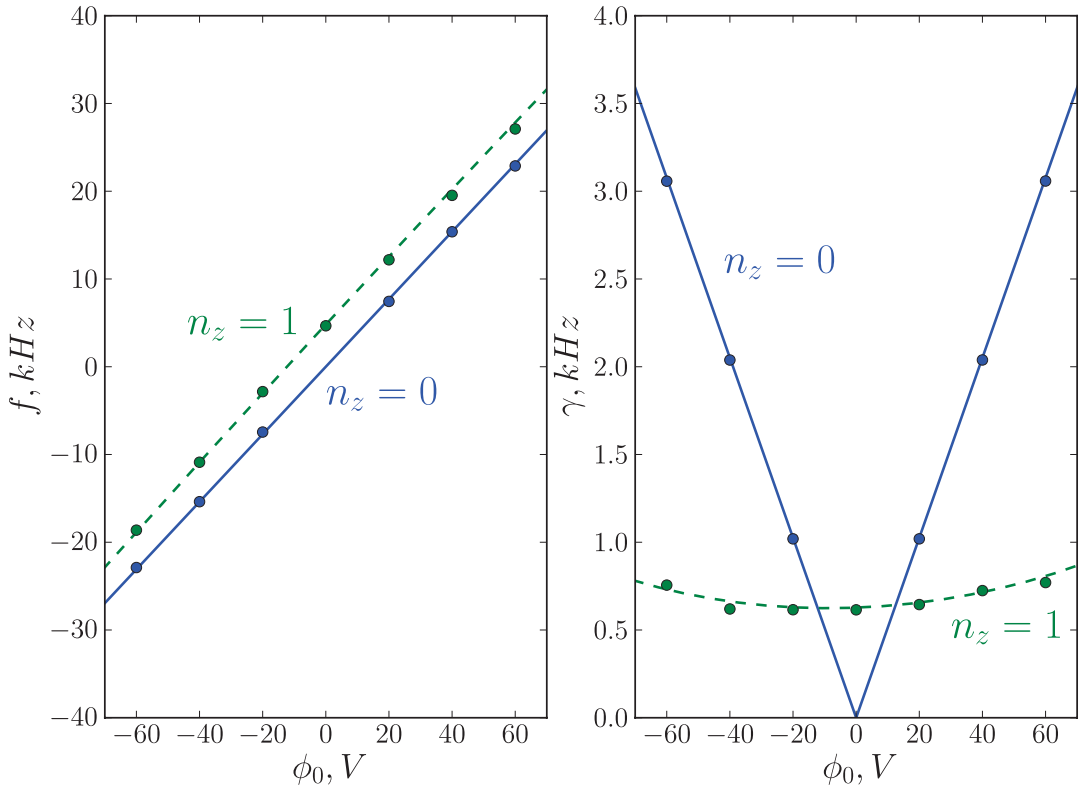}
\end{center}
\caption{\label{figIC_scan} (Color Online) Interchange mode ($k_\para=0$)
    destabilized by uniform rotation and drift-interchange mode
    ($k_\para=2\pi/L$).}
\end{figure}

The system of time-evolution equations used in BOUT to reproduce the
interchange mode can be obtained from Eqs.~(\ref{eqBN}-\ref{eqVort})
by setting $\kpar=0$:
\beqar
\label{eqIC_N}
\pdt N & = & - \vE\cdot\grad N\\
\label{eqIC_Vort}
\pdt\varpi & = &
-\vE\cdot\grad\varpi 
+ \bvec\times\grad N\cdot\grad{v_E^2}/2
-\nuin\varpi
\eeqar
where all variables ($N$, $\vE$, $\varpi$) contain both the
equilibrium part and the fluctuating component. These equations are
linearized in BOUT, and solved with the same parameters as used in the
two analytic examples discussed above
(Fig.~\ref{figIC_analytic}). BOUT simulation recovers the interchange
mode solution for both limits (slowly varying exponential and
piecewise constant profiles) which verifies the correct implementation
of the new terms involving background flows in BOUT.

In order to investigate the effect of uniform rotation on the
interchange and drift-interchange instabilities in LAPD plasmas, a
configuration with experimental density profile and $\phi_0(r)\sim
r^2$ is considered. The results of this calculation are presented in
Fig.~(\ref{figIC_scan}) as a function of rotation velocity. Two axial
harmonics are shown: $n_z=0$ (pure interchange mode) and $n_z=1$
(drift-interchange instability). At $\phi_0=0V$, $n_z=1$ branch
corresponds to a pure drift mode. As $\phi_0$ increases, the frequency
of this mode is Doppler-shifted and the growth rate is modified by the
centrifugal force. At large rotation velocities, the disparity between
the large real part of the frequency and the small growth rate is hard
to resolve numerically using an initial-value code, so the BOUT results
slightly deviate from the direct eigenvalue solution of the dispersion
relation.


\section{Linear instabilities in LAPD}\label{secDW_KH_IC}

Now that simple analytic solutions for each of the instabilities
supported by Eqs.~(\ref{eqfN}-\ref{eqfdivJ}) have been presented, 
linear instabilities for LAPD parameters and experimental profiles
will be considered and the growth rates for the different mode
branches will be compared.

\begin{figure}[htbp]
\begin{center}
\includegraphics[width=0.9\textwidth]{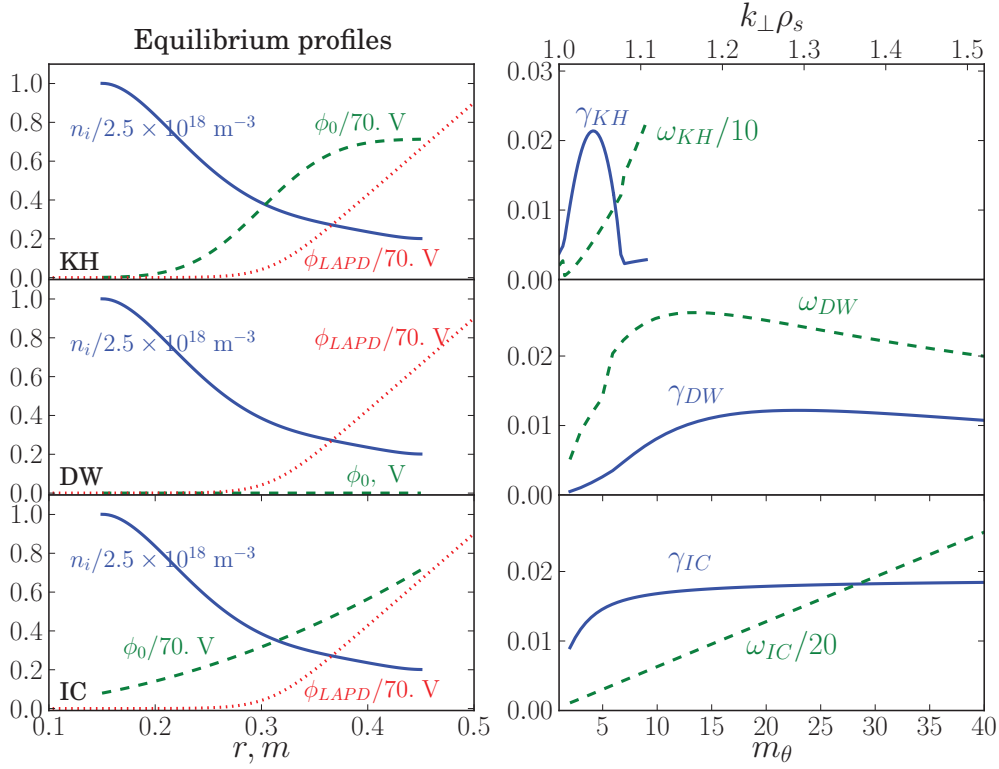}
\end{center}
\caption{\label{figKH_DW_IC_scan} (Color Online) Kelvin-Helmholtz, drift and
interchange branches of the dispersion relation for LAPD parameters as
a function of azimuthal mode number. Left: equilibrium profiles,
right: frequency and growth rate of the instability. For KH and IC
$n_z$=0, for DW $n_z$=0.5. For DW case $\phi_0(r)=0$, for IC
instability $\phi_0(r)\sim r^2$ (uniform rotation), for KH mode
$\phi_0(r)$ is given by Eq.~(\ref{eqPhi0Gauss}). As a reference, the
experimentally measured $\phi_{LAPD}$ profile is shown in dotted line (left).}
\end{figure}

Fig.~(\ref{figKH_DW_IC_scan}) shows the growth rates and frequencies
of the KH, drift and interchange modes for LAPD parameters using
experimentally measured density profiles. The complete set of
parameter values including the polynomial fit of the experimental
density profile is shown in Appendix~\ref{secAppendixProfiles}. Three
different model background potential profiles are chosen here to
separate the instability branches: same profile as used in
Fig.~\ref{figKHgauss} for the KH mode (given by
Eq.~(\ref{eqPhi0Gauss})), uniform rotation profile $\phi_0(r)\sim r^2$
for the interchange mode, and zero potential for the drift wave
instability. The magnitude of the radial potential drop in the
KH and IC cases is of the same order as the measured value in biased
discharge experiments~\cite{Maggs2007}. Even though a direct
comparison of the three solutions is not possible because the
background flow profiles and axial mode numbers are not the same, it
is still informative to note that the growth rates of all three
branches of instability are of similar magnitude. Therefore, all three
instabilities can potentially compete in LAPD plasmas.

\begin{figure}[htbp]
\centerline{\includegraphics[width=0.9\textwidth]{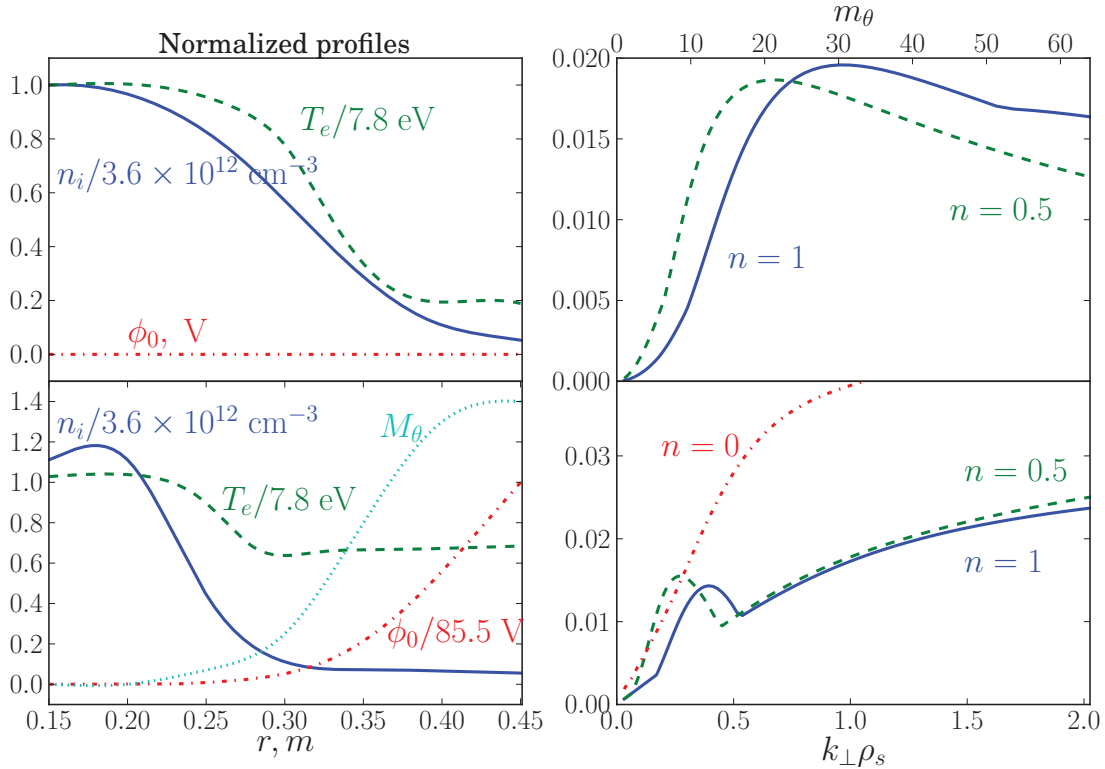}}
\caption{\label{figLAPD_wbias} (Color Online) Equilibrium profiles of the
density, electron temperature, potential and Mach number (left) and
the fastest growth rates (right) of the perturbation with axial mode number
$n_z=0,0.5,1$. Top: unbiased LAPD discharge. Bottom: LAPD discharge with
applied radial bias. Experimental data taken from Maggs et al.\cite{Maggs2007}.  }
\end{figure}

\begin{figure}[htbp]
\centerline{\includegraphics[width=0.7\textwidth]{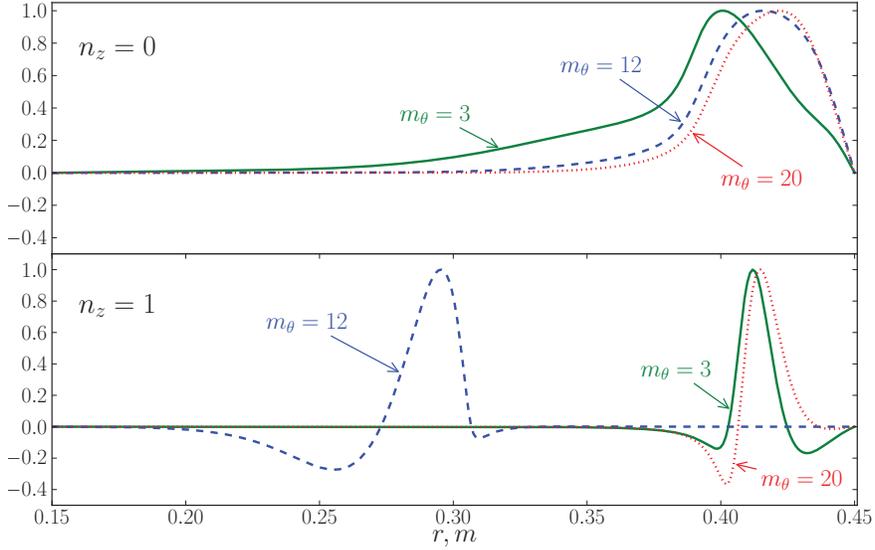}}
\caption{\label{figLAPD_eig} (Color Online) Eigenfunctions of the perturbed
potential in the biased plasma configuration (case shown in
Fig.~\ref{figLAPD_wbias}, bottom) for axial mode number $n_z=0,1$,
azimuthal harmonics $m_\theta=3,12,20$.}
\end{figure}

Similar results are observed in a calculation of linear growth rates
when using self-consistent, experimentally measured profiles of
density, electron temperature and flow (Fig.~\ref{figLAPD_wbias}). Two
cases are considered here, biased and unbiased plasma discharges (with
and without bias-driven azimuthal edge flow)~\cite{Maggs2007}. In the
unbiased configuration, the azimuthal flow values are much smaller
than in the biased case, so we use zero azimuthal flow for this
calculation. In the unbiased case (Fig.~\ref{figLAPD_wbias}, top),
only the drift wave branch is present, with comparable maximum growth
rates for $n_z=1$ and $n_z=0.5$.   In the biased case (Fig.~\ref{figLAPD_wbias},
bottom), the growth rates at $m_\theta\lesssim 10$ for the three
harmonics $n_z=0$, $n_z=0.5$ and $n_z=1$ are comparable. From the
eigenfunction analysis, it can be concluded that $n_z=1$ harmonic is
predominantly interchange at $m_\theta\lesssim 5$, then drift
wave-like at $5\lesssim m_\theta\lesssim 17$ and again IC-like at
higher $m_\theta$. An example of the eigenfunctions of the potential
perturbation for the biased case is shown in
Fig.~\ref{figLAPD_eig}. At $m_\theta=3$ and $m_\theta=20$, the axial
mode $n_z=1$ is localized near the edge of the plasma where the
azimuthal flows are strongest (see $M_\theta$ profile in
Fig.~\ref{figLAPD_wbias}, bottom), which is consistent with the
rotational interchange instability. At $m_\theta=12$, the $n_z=1$
harmonic eigenfunction is localized near
$r\sim 28$ cm, where the gradients of the density and electron
temperature are strongest, which indicates the drift-wave-like character of
the mode.

The real frequencies of these modes are consistent with experimental
observation; in the unbiased case, at the peak of the growth rate for
$n_z=0.5$ ($m \sim 20$) the mode frequency is $f = 4.7$kHz which is in
the heart of the measured broadband fluctuation spectrum in unbiased
plasmas, although a lower mode number would be
consistent with the measured correlation function~\cite{Carter2009}.  
In the biased case, the local maximum of the
growth rate of $n_z=0.5$ mode is at $m_\theta=8$, which is also
consistent with measured LAPD value of $m_\theta\lesssim
10$~\cite{Carter2009}. Higher growth rates at large $m_\theta$ might
not be relevant when viscosity effects are included in the
calculation, since high $k_\perp$ modes will be damped by viscosity.
The computed linear eigenfunctions are consistent with the observed
fluctuation profiles in the unbiased case, localized to the density
gradient region.  In the biased case, eigenfunctions localized to the
region of strong density gradient are found as well as flow-driven
modes that are localized to the far edge away from the strong gradient
region.  The latter is consistent with the observation of increased
electric field fluctuations in the far-edge plasma with increased bias
(see Fig. 9c of Ref. 10).  The linear prediction that Kelvin-Helmholtz
and/or rotational interchange might be the dominant instabilities in
the biased case could also be consistent with measurements of the cross
phase between density and electric field fluctuations. In going from
unbiased to biased plasmas in LAPD, a dramatic change in the cross
phase is observed, which could be consistent with a change in the
dominant instability~\cite{Carter2009}.

A detailed comparison with the experimental data requires
nonlinear analysis and simulation, which is the subject of a companion
paper~\cite{Popovich2010b}. However, it is still illustrative to apply quasilinear
theory or mixing length arguments~\cite{Wesson2004,Kadomtsev1965}
using the linear calculation results presented above. For drift waves driven by the
background density gradient the mixing length estimate assumes that
the saturation is reached when the perturbed gradients become
comparable to the equilibrium gradients:
\beq
n\kperp \sim n_0/L_n
\eeq
so that
\beq
n/n_0 \sim q\phi/T_e \sim 1/\kperp L_n
\eeq
The weak turbulence theory modifies this estimate by a factor $\sqrt{\gamma/\omega_*}$:
\beq
n/n_0 \sim q\phi/T_e \sim \sqrt{\gamma/\omega_*}/\kperp L_n
\eeq
where $\gamma$ and $\omega_*$ are the growth rate and frequency of the
fastest linear mode in the system.

In the unbiased case (Fig.~\ref{figLAPD_wbias}, top), the maximum
growth rate is achieved for modes with $m_\theta\sim 20-30$, so
$k_\perp\sim m_\theta/r\sim 100{\rm~m^{-1}}$. The frequency is close
to the growth rate for these modes, so $\gamma/\omega_*\sim 0.5-1$. The
background density gradient scale length near the radial location of
interest (cathode edge at $\sim 28~\cm$ where the equilibrium density
gradient is mostly localized) is $L_n=n_0/n_0'\sim
0.1{\rm~m}$. Therefore, both the simple mixing length argument and the
weak turbulence theory give a similar estimate for the saturated level
of turbulence, $n/n_0 \sim q\phi/T_e \sim 10\%$. This estimate is
close to the observed amplitude of fluctuations in LAPD measurements
and in the nonlinear simulations of LAPD
discharge~\cite{Carter2009,Popovich2010b}. The diffusion coefficient
estimate based on the mixing length argument,
$D\sim\gamma/\kperp^2\sim 2{\rm~m^2/s}$, is close to the value
calculated from a saturated state in a self-consistent nonlinear
simulation $D\sim 3{\rm~m^2/s}$~\cite{Popovich2010b}. This value is comparable to Bohm
diffusion, $D_B\sim 8{\rm~m^2/s}$, and diffusive transport with a Bohm
diffusion coefficient has been found to describe the measured profiles
well in the unbiased case~\cite{Maggs2007}.

The biased configuration has reduced radial transport due to strong
azimuthal flows and is better described by classical diffusion
coefficient~\cite{Maggs2007}. Detailed analysis of this case requires
a self-consistent nonlinear simulation that takes into account the
average radial electric field profile; this will be the subject of
future work.



\section{Effect of ion-neutral collisions}\label{secVerif_NU}

\begin{figure}[htbp]
\begin{center}
\includegraphics[width=0.9\textwidth]{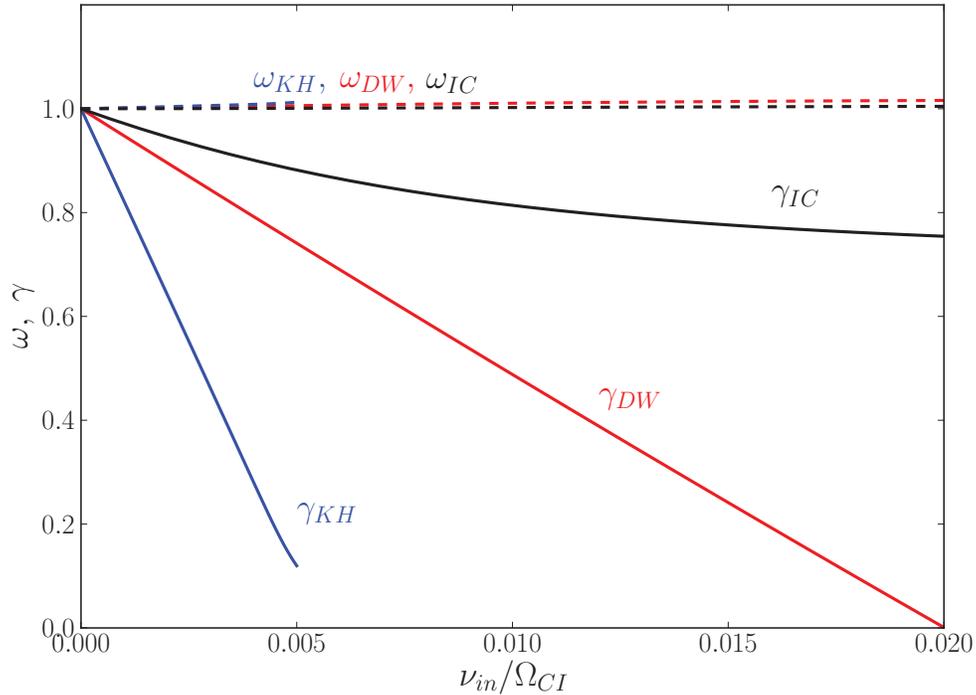}
\end{center}
\caption{\label{fignuin} (Color Online) Effect of ion-neutral collisions on the
    Kelvin--Helmholtz (KH), drift wave (DW) and interchange (IC)
    branches of the dispersion relation. All scans are normalized to
    the corresponding value at $\nuin=0$. Solid line -- growth rate,
    dashed -- frequency. Estimate for typical LAPD parameters:
    $\nuin/\wci\sim\enum{2}{-3}$.}
\end{figure}

The results presented in previous sections do not include ion-neutral
collision terms that enter the vorticity equation
Eq.~(\ref{eqVort}). The general effect of the $\nuin$ term is to damp
the vorticity perturbations (as can be seen from Eq.~(\ref{eqVort}))
and to stabilize the wave. In Fig.~\ref{fignuin}, variation of the
real frequency and growth rates for the three modes (drift, KH, and
IC) is shown as a function of the ion-neutral
collisionality parameter $\nuin$. Each branch is taken at a fixed
azimuthal mode number $m_{\theta}$ that corresponds to the maximum
growth rate without neutrals (same solution as in
Fig.~\ref{figKH_DW_IC_scan}), except for the interchange branch, where
$m_\theta=10$ is chosen). All of the frequencies and growth rates are
normalized to the corresponding values at $\nuin=0$.

When ion-neutral collisions are included, the drift wave growth rate
decreases and the mode can be completely stabilized at sufficiently
high neutral density. For a typical LAPD discharge, the rough estimate
of $n_N$ is $\sim5\times10^{11}{\rm~cm}^{-3}~~$~\cite{Maggs2007} which
translates into $\nuin \sim 2\times10^{-3}\wci$. At these values of
$n_N$, the effect of the neutrals on the linear stability is
relatively weak. To completely stabilize the drift mode, $n_N$ should
be larger by a factor of 10 (Fig.~\ref{fignuin}, red). However, due to
significant uncertainty in the values of neutral density in LAPD,
ion-neutral collisions can potentially be important.  More
importantly, initial nonlinear simulations using BOUT show that
even at the values near the estimated
$\nuin\sim2\times10^{-3}\wci$, the neutral damping is important for the
dynamics of the self-generated zonal flows~\cite{Popovich2010b}.

Compared to the drift mode, the KH instability is more strongly
affected by the ion-neutral collisions. Compared to the neutral-free
case, at the estimated for LAPD level of ion-neutral collisions, the
growth rate drops by $\sim$35\% and the mode is completely stabilized
at $\nuin/\wci\sim 0.006$.

The interchange mode turns out to be weakly affected by ion-neutral
collisions. For all three instability branches, the frequency of the
mode remains nearly constant in the range of relevant values of
neutral collisionality (Fig.~\ref{fignuin}, dashed lines).


\section{Conclusions}\label{secCon}

The 3D initial value fluid code for
tokamak edge plasma has been adapted to LAPD geometry. A separate
eigenvalue solver for BOUT set of linearized equations has been
developed for an independent verification of BOUT results when an
analytic solution is not available. Background flow terms have been
added to BOUT equations to allow simulation of flow-driven
instabilities. Periodic boundary conditions has been adopted in the
parallel direction as a first step. A more realistic model of sheath
boundary conditions will be implemented in future simulations to
capture the effect of the parallel boundary on the dynamics of the
average radial electric field.

Starting from a system of 3D plasma fluid equations, the derivation of
a dispersion relation is presented that includes three plasma
instability branches: resistive drift mode, Kelvin-Helmholtz mode, and
interchange mode; the latter two driven by plasma azimuthal flow. 
It is demonstrated that for LAPD parameters the growth rates for all
three branches may be comparable, so all three physical mechanisms are
potentially important. Interaction with neutrals, for the estimated
LAPD neutral density, does not significantly affect the linear
stability of considered modes. However, neutral dynamics can be
important for the zonal flow generation in nonlinear simulations.

The initial value solution obtained with BOUT accurately reproduces
analytic calculations of the properties of the three instabilities,
including growth rates, frequencies and eigenfunctions. The code
solution is in full agreement with analytic and eigenvalue solutions,
for both model profiles and experimentally relevant profiles, which
lends confidence for proceeding with nonlinear simulations and
validation of BOUT against LAPD measurements. Aspects of these linear
theoretical estimates (dominant mode numbers, mode frequency, and
quasilinear estimates of fluctuation amplitudes and diffusion
coefficient) are consistent with the experimental measurements in
LAPD. However, more detailed comparison with experiment requires
self-consistent nonlinear simulations and this work is
underway. Initial nonlinear calculations based on the model discussed
in this work, and detailed comparisons with experimental data, will be
presented in a companion paper~\cite{Popovich2010b}.


\begin{acknowledgments}

This work was supported by DOE Fusion Science Center Cooperative
Agreement DE-FC02-04ER54785, NSF Grant PHY-0903913, and by LLNL under
DOE Contract DE-AC52-07NA27344.  BF acknowledges support through
appointment to the Fusion Energy Sciences Fellowship Program
administered by Oak Ridge Institute for Science and Education under a
contract between the U.S. Department of Energy and the Oak Ridge
Associated Universities.

\end{acknowledgments}

\appendix

\section{Derivation of fluid equations}\label{secAppendixFluid}

The perpendicular component of the current in Eq.~(\ref{eqfdivJ}) is
found from the fluid equation for the ions Eq.~(\ref{eqfvpari}).  Note
that the viscosity tensor $\Pi$ and ion pressure terms are dropped
here, since we neglect the ion temperature effects in this
work. Solving it for ion velocity $\vvi$ in the Lorentz term, we
obtain $\vect{v}_{i\perp}$ as a sum of the $\Evec\times\Bvec$,
polarization and Pedersen drifts:
\beqar 
\vect{v}_{i\perp} & = & \vvE + \vect{v}_{pi} + \vect{v}_{fi}
\eeqar
where $\vvE = c\Evec\times\Bvec/B^2$, $\vect{v}_{pi} = 
\displaystyle{\frac{1}{\wci}\left(\partial_t + 
\vvi\cdot\grad\right)\vvi}$, $\vect{v}_{fi} = 
\displaystyle{\frac{\nuin}{\wci}\Bvec\times\vvi/B}$. 

The main contributions to the perpendicular part of the current
divergence come from the ion polarization current (the electron
olarization drift is smaller by mass ratio) and Pedersen current:
\beqar
\nn\div\Jvec_\perp \approx \div\left(en\vect{v}_{pi} + en\vect{v}_{fi}\right)
& &\\
\nn & & \hspace{-2cm}= \frac{1}{\wci}\div \left\{en\bvec\times\left(
\pdiff{\vect{v}_i}{t} + \vect{v}_i\cdot\grad\vect{v}_i
\right) + en\nuin\bvec\times\vvi \right\}\\
\nn& & \hspace{-2cm}\approx \frac{1}{\wci}\div \left\{en\bvec\times\left(
\partial_t + \vvE\cdot\grad\right)\vvE + en\nuin\bvec\times\vvE\right\}
\eeqar

To make the linear expansion of the current continuity equation
exactly equivalent to the linearized BOUT vorticity equation discussed
below, we employ the approximation $\div(n\vvi)\sim\vvE\cdot\grad n$
(well satisfied for typical LAPD parameters), the same way it is done
in previous work by Simakov and Catto~\cite{Simakov2003} (Eq. D3):
\beqar
\div\Jvec_\perp
& \approx & -\frac{m_ic^2}{B^2}\div \left\{\left(\partial_t + \vvE\cdot\grad\right)(n\gradperp\phi) + \nuin n\gradperp\phi\right\}
\eeqar
Substituting this expression in the charge conservation equation
Eq.~(\ref{eqfdivJ}), combining with the continuity equation
Eq.~(\ref{eqfN}) and parallel projection of the electron momentum
equation Eq.~(\ref{eqfvpar}) and linearizing, we obtain:
\beqar\label{eqs_origa}
\nn
\partial_t N +
\bo\times\gradperp\phi_0\cdot\grad N & = & -
\bo\times\gradperp\phi\cdot\grad N_0 - N_0\gradpar\vpe\\
\partial_t{\vpe}+\bo\times\gradperp\phi_0\cdot\grad\vpe & = &
-\mu\frac{T_{e0}}{N_0}\gradpar N + \mu\gradpar\phi - \nue\vpe\\
\nn
N_0\gradpar\vpe & = & -\gradperp\cdot
       \left(
             N_0\partial_t\gradperp\phi + \partial_t N\gradperp\phi_0\right.\\
\nn& &\hspace{2.0cm} + \bo\times\gradperp\phi_0\cdot\grad\left(N_0\gradperp\phi_0\right)\\
\nn
& &\hspace{2.0cm} + \bo\times\gradperp\phi_0\cdot\grad\left(N_0\gradperp\phi\right)\vphantom{\frac{}{1}}\\
\nn
& &\hspace{2.0cm} + \bo\times\gradperp\phi\cdot\grad\left(N_0\gradperp\phi_0\right)\vphantom{\frac{}{1}}\\
\nn
& &\hspace{2.0cm} + \bo\times\gradperp\phi_0\cdot\grad\left(N\gradperp\phi_0\right)\vphantom{\frac{}{1}}\\
\nn
& &               + N_0\nuin\gradperp\phi_0
                  + N_0\nuin\gradperp\phi
                  + N\nuin\gradperp\phi_0
\left. \vphantom{\pdiff{1}{t}}\right) 
\eeqar
We project these equations on cylindrical coordinates $(r,\theta,z)$
and assume the fluctuations are of the form
$f(\vect{x})=f(r)\exp(im_\theta\theta + i\kpar z - i\omega t)$. Solving the
first two equations for $N$ and $\vpe$, and substituting them in the
current equation, we obtain 1D equation for the perturbed potential:
\beq\label{eqPhi1Da}
C_2(r)\phi'' + C_1(r)\phi' +C_0(r)\phi = 0,
\eeq
\beqar
C_2(r) & = & \left(\nuin - i\tomega\right)\\
C_1(r) & = & \left(\nuin - i\tomega\right)
             \left(\frac{1}{r} - \frac{1}{L_n} + \phi_0'\lambda_N\right)
            +im_\theta\frac{1}{rL_n}\phi_0'\\
C_0(r) & = & \left(\nuin - i\tomega\right)
                   \left(
                   - \frac{m_\theta^2}{r^2} 
                   + \lambda_N\phi_0'\left(\frac{1}{r}-\frac{1}{L_n}\right)
                   + (\lambda_N\phi_0')'
                   \right)\\
         & & + \frac{im_\theta}{r^3}
                 \left(
                 \phi_0' - r\phi_0'' - r^2\phi_0'''
                 -\frac{r}{N_0}(rN_0'\phi_0')' + \frac{r^2}{L_n}\phi_0''
                 \right)\\
         & & + ik_\parallel\lambda_v + im_\theta\frac{1}{r}\lambda_N\phi_0'\phi_0'',
\eeqar
where
\beqar
\nn \lambda_v(r,\tomega) & = &
i\kpar\mu \frac{1 - \frac{T_{e0}}{\tomega L_n}\frac{m_\theta}{r}}
{\nue - i\tomega + i\kpar^2\mu\frac{T_{e0}}{\tomega}}\\
\nn \lambda_N(r,\tomega) & = &
\frac{i\kpar^2\mu + \frac{m_\theta}{r}\frac{1}{L_n}\left(\nue-i\tomega\right)}
{\tomega\left(\nue-i\tomega\right) + i\kpar^2\mu T_{e0}}\\
\nn L_n & = &-\frac{N_0}{N_0'}, ~~~~~~ \tomega=\omega - \frac{m_\theta}{r}\phi_0'\\
\eeqar

\section{Derivation of the vorticity equation}\label{secAppendixVort}

Expanding the charge conservation equation $\div\Jvec = 0$ as
described in section \ref{secEquations}, we can write
\beq
0=\div\Jvec_{||} + \div\Jvec_\perp = 
\gradpar(N\vpar)
-\gradperp\cdot \left\{\left(\partial_t + \vvE\cdot\grad\right)(n\gradperp\phi) + \nuin n\gradperp\phi\right\}
\eeq
Introducing the potential vorticity defined as $\displaystyle{\varpi\defeq\gradperp\cdot\left(N\gradperp\phi\right)}$, we can rewrite the second term:
\beqar\label{eqAppSimpl1}
\nn
& &\hspace{-1.5cm} -\gradperp\cdot \left\{\left(\partial_t + \vvE\cdot\grad\right)(N\gradperp\phi) + \nuin n\gradperp\phi\right\}\\
\nn & &\hspace{-1cm} = -\partial_t\varpi - \vvE\cdot\grad\varpi 
- \gradperp \vvE : \gradperp(N\gradperp\phi) - \nuin\varpi\\
& &\hspace{-1cm} = -\partial_t\varpi - \vvE\cdot\grad\varpi 
- \gradperp\vvE:\gradperp N\gradperp\phi - N\gradperp\vvE:\gradperp\gradperp\phi 
- \nuin\varpi
\eeqar
The fourth term in this expression vanishes:
\beqar
\gradperp\vvE:\gradperp\gradperp\phi
& = &
\gradperp(\bxgp\phi):\gradperp\gradperp\phi\\
& &\hspace{-4cm}=
\nn\frac{1}{2}\left(
\gradperp^2(\gradperp\phi\cdot\bxgp\phi) 
- (\gradperp^2\gradperp\phi)\cdot(\bxgp\phi)
- \gradperp^2(\bxgp\phi)\cdot\gradperp\phi
\right)\\
& &\hspace{-4cm}=
\nn\frac{1}{2}\left(
- (\gradperp^2\gradperp\phi)\cdot(\bxgp\phi)
- (\bxgp\gradperp^2\phi)\cdot\gradperp\phi
\right) = 0
\eeqar
The third term in Eq.~(\ref{eqAppSimpl1}) can be simplified as follows:
\beqar
\gradperp\vvE:\gradperp N\gradperp\phi
& = &
\nn\left\{\gradperp\phi\cdot\gradperp(\bxgp\phi)\right\}\cdot\gradperp N\\
& = &
\nn(\gradperp N\times\bvec)\cdot(\gradperp\phi\cdot\gradperp\gradperp\phi)\\
& = &
\nn\frac{1}{2}(\grad N\times\bvec)\cdot(\gradperp\gradperp\phi^2)
=\frac{1}{2}(\grad N\times\bvec)\cdot\gradperp\vvE^2\\
\eeqar

Collecting all terms, we can write the equation for the evolution of
potential vorticity:
\beqar\label{eqAppVort}
\nn
\partial_t\varpi = 
- \vvE\cdot\grad\varpi 
+ \gradpar(N\vpar)
+ \frac{1}{2}(\bvec\times\grad N)\cdot\gradperp\vvE^2
- \nuin\varpi
\eeqar

\section{Parameters and profiles for the benchmark case}\label{secAppendixProfiles}

Parameters and profiles used for the simulation are presented in Fig.~\ref{figKH_DW_IC_scan}.

\vspace{1cm}

Common parameters for all 3 cases (drift wave, Kelvin-Helmholtz, interchange):
\medskip 

Helium plasma, once ionized $Z=1$

Radial interval
$r_a\leq r \leq r_b$,  $r_a=0.15~{\rm m}$, $r_b=0.45~{\rm m}$

$B_0=0.04~\T$, $T_e=5~\eV$, $\nuin=0$, $L_z=17~{\rm m}$

Density profile is a polynomial fit to the experimental profile
$n_i(r) = n_0\sum_{i=0}^{5}c_i r^i$, 

$\{c_i\}= \{-5.4638, 124.624, -882.24, 2863.636, -4436.36, 2666.664\}$,
$n_0=2.5\times10^{18}~{\rm m}^{-3}$.

\vspace{1cm}

Different parameters for each of the 3 cases:

\medskip 
Drift wave case:
$n_z=0.5$, $\phi_0(r)=0$.

\medskip 
Kelvin-Helmholtz case:
$n_z=0$, 
\beqar
\label{eqPhi0Gauss}
\phi_0(r)&=&\phi_x\left(F(x-1)+F(x+1)-2F(x)\right),\\ 
\nn F(x)&=&\frac{1}{8}\left(\frac{2wx}{\sqrt{\pi}}e^{-\frac{x^2}{w^2}} 
+ (w^2+2x^2){\rm erf}\left(\frac{x}{w}\right) \right),
\eeqar

$x=4(r-r_a)/(r_b-r_a)-2$, $w=0.8$, $\phi_x=50~V$.

\medskip 
Interchange case:
$n_z=0$, 
$\displaystyle{\phi_0(r)=\phi_x\left(\frac{r}{r_b}\right)^2}$, $\phi_x=50~V$.

\medskip 
Boundary conditions: periodic in the azimuthal and axial directions; $\phi(r_a)=\phi(r_b)=0$ radially.


%
%
%
\end{document}